\documentclass[twocolumn,preprintnumbers,nofootinbib,prl,showpacs]{revtex4}

\usepackage{graphicx}
\usepackage{graphics}
\usepackage{dsfont}
\usepackage{epsfig}
\usepackage{amsmath,amssymb,amsthm,amscd}

\usepackage{hyperref}
\newcommand{\beq}{\begin{equation}}
\newcommand{\eeq}{\end{equation}}
\newcommand{\be}{B_\oplus}

\newcommand{\Tr}{\textrm{Tr}}

\def\ket{\rangle}
\def\bra{\langle}

\def\be{\begin{equation}}
\def\ee{\end{equation}}
\def\baray{\begin{eqnarray}}
\def\earay{\end{eqnarray}}
\def\bes{\begin{eqnarray}}
\def\ens{\end{eqnarray}}
\def\ba{\begin{eqnarray}}
\def\ea{\end{eqnarray}}

\def\bes {\begin{eqnarray}}
\def \ens {\end{eqnarray}}

\begin{document}

\newcommand{\bea}{\begin{eqnarray}}
\newcommand{\eea}{\end{eqnarray}}
\newcommand{\barr}{\begin{array}}
\newcommand{\earr}{\end{array}}

\pagestyle{plain}

\preprint{USTC-ICTS-19-27}

\title{Entropy of Berenstein-Maldacena-Nastase Strings}

\author{Min-xin Huang\footnote{email: minxin@ustc.edu.cn} 
}

\affiliation{Interdisciplinary Center for Theoretical Study, School of Physical Sciences, \\     University of Science and Technology of China,  Hefei, Anhui 230026, China,  \\
and Peng Huanwu Center for Fundamental Theory, Hefei, Anhui 230026, China }


\begin{abstract}
In a previous paper, we proposed a probability interpretation for higher genus amplitudes of BMN (Berenstein-Maldacena-Nastase) strings in a pp-wave background with infinite negative curvature. This provides a natural definition of the entropy of a BMN string as the Shannon entropy of its corresponding probability distribution. We prove a universal upper bound that the entropy grows at most logarithmically in the strong string coupling limit. We also study the entropy by numerical methods and discuss some interesting salient features.

\end{abstract}
\maketitle

\section{Introduction} 

The holographic AdS/CFT correspondence \cite{Maldacena, Gubser, Witten} has been very influential  in recent decades, due to its wide-ranging applications in various topics in theoretical physics. In our previous works \cite{Huang:2002, Huang:2010, Huang:2019}, we have accomplished some spectacular quantitative tests of the AdS/CFT correspondence in a stringy regime. In a pp-wave background with infinite negative curvature, stringy states are identified with the Berenstein-Maldacena-Nastase (BMN) operators \cite{BMN} in the free $\mathcal{N}=4$ $SU(N)$ super-Yang-Mills theory. In this case, string theory turns out to be also extremely simplified. We proposed the string (loop) amplitudes can be computed by cubic diagrams without propagators, and are related to the field theory side calculations by the so-called ``factorization principle." The usually notoriously difficult problem of computing higher loop string amplitudes with stringy excited modes, corresponding to nonplanar BMN correlators here, becomes a straightforward exercise in our case and can be computed explicitly on both sides of the correspondence. 

 Let us introduce some notations. The BMN operators up to two string oscillator modes, orthonormal at planar level, are the following:
 \begin{eqnarray} \label{BMNoperators}
&& O^{J}  = \frac{1}{\sqrt{JN^J}}TrZ^J,  ~~~~
O^{J}_{0} = \frac{1}{\sqrt{N^{J+1}}} Tr(\phi^{I} Z^{J}), \nonumber \\
&& O^J_{-m,m} = \frac1{\sqrt{JN^{J+2}}} \sum_{l=0}^{J-1}e^{\frac{2\pi iml}{J}} 
Tr(\phi^{I_1} Z^l\phi^{I_2} Z^{J-l}). \nonumber
\end{eqnarray}
Here we take the BMN limit $J, N\rightarrow \infty$, and identify $g:=\frac{J^2}{N}$ as the effective string coupling constant. Some higher genus correlators were first computed by early papers on the subject \cite{KPSS, Constable1}. For example, the free torus (genus one) two-point function is computed by dividing the long strings into four segments and Wick contracting according to a ``short process" $(1234)\rightarrow (2143)$. The result for BMN operators with two stringy modes can be written as an integral and evaluated: 
\begin{eqnarray} \label{torusintegral}
&& \bra \bar{O}_{-m,m}^J O_{-n,n}^J \ket_{\textrm{torus}}   \nonumber \\ 
&=& g^2 \int_{0}^1 d x_1 dx_2 dx_3 dx_4 \delta(x_1+x_2+x_3+x_4-1)   \nonumber \\
&& [\int_0^{x_1} dy_1 e^{-2\pi i (m-n) y_1}]   
 \cdot  [\int_0^{x_1} dy_2 e^{2\pi i (m-n) y_2} \nonumber \\ && +  e^{2\pi im(x_3+x_4) } \int_{x_1}^{x_1+x_2} dy_2 e^{2\pi i (m-n) y_2} 
\nonumber \\ && +e^{2\pi i m(x_4-x_2) } \int_{x_1+x_2}^{1-x_4} dy_2 e^{2\pi i (m-n) y_2} 
\nonumber \\ && +e^{-2\pi i m(x_2+x_3) } \int_{1-x_4}^{1} dy_2 e^{2\pi i (m-n) y_2} ]  \\
&=& \left\{
\begin{array}{cl}
\frac{g^2}{24},    &   m=n=0;   \\
0,              &  m=0, n\neq0,   \\
&  \textrm{or}~n=0, m\neq0; \\
g^2(\frac{1}{60} - \frac{1}{24 \pi^2 m^2} + \frac{7}{16 \pi^4 m^4}),   &  m=n\neq0; \\
\frac{g^2}{16\pi^2m^2} ( \frac{1}{3}+\frac{35}{8\pi^2m^2}),  &  m=-n\neq0;    \nonumber \\
\frac{g^2}{4\pi ^{2}(m-n)^2} ( \frac{1}{3}+\frac{1}{\pi
^2n^2}+\frac{1}{\pi ^2m^2}    \nonumber \\ -\frac{3}{2\pi ^2mn}-\frac{1}{2\pi
^2(m-n)^2}) & \textrm{all~other~cases.} 
\end{array}
\right.
\end{eqnarray}
Here as in our previous papers we only consider free gauge theory, and omit the universal spacetime dependent factor in correlators. 

In our recent paper \cite{Huang:2019} we proposed a physical probability interpretation of the free higher genus amplitudes of two single string BMN states. We proposed that the probability of preparing a BMN string $O_{-m,m}$, then observing another BMN string $O_{-n,n}$, can be written as
\begin{eqnarray} \label{pmatrix}
p_{m,n} = \frac{g}{2\sinh(g/2) } \sum_{h=0}^{\infty} \bra \bar{O}_{-m,m}^J O_{-n,n}^J \ket_{h},
\end{eqnarray}
with the sum over genus $h$. The supporting evidence of our proposal is that each term in the formula is always non-negative $ \bra \bar{O}_{-m,m}^J O_{-n,n}^J \ket_{h} \geq 0$, and it is simply normalized by the vacuum correlator to sum over all final states to unity $\sum_{n=-\infty}^{\infty} p_{m,n} =1$ for any initial mode $m$. 

This provides an interesting new entry of holographic dictionary beyond supergravity. Denote $|n\ket$ the orthonormal BMN states of free string theory and assume the string interactions are described by a unitary operator $\hat{U}(g)$, our proposal implies the matrix element $p_{m,n}$ of the two point function does not naively correspond to the quantum transition amplitude $\bra m| \hat{U}(g) |n\ket$, but rather to its norm square 
\begin{equation} \label{newentry}
p_{m,n} = |\bra m| \hat{U}(g) |n\ket|^2. 
\end{equation}
This is further supported by unitarity arguments. It is interesting to note that the cubic string vertex is ``virtual" since they are vanishing in the strict BMN limit, but still useful because infinitely many of them can combine to make a finite contribution.  In this sense the single strings form a complete Hilbert space by themselves $\sum_n |n\ket \bra n|=1$, the multistrings are regarded as virtual states. For more details see \cite{Huang:2019}.  

We note that the spacetime is highly compressed due to the infinite negative curvature, so its coordinates do not appear in the transition amplitudes. It is helpful to compare with the scattering amplitude in quantum field theory in flat spacetime. We usually quantize the free field theory to obtain an orthogonal basis of the Hilbert space, while the interaction is described separately by a unitary operator depending on the coupling. It would be extremely awkward to directly quantize an interacting quantum field theory or string theory, and we do not know such an example. The amplitude $\bra m| \hat{U}(g) |n\ket$ analogously plays the role of the S-matrix of an initial state becoming asymptotically a final state, without the details of time evolution.

It is well known that the string perturbative series is usually asymptotically double-factorially divergent. However, in our case, since each term in (\ref{pmatrix}) is non-negative, it is apparent that our perturbative series is actually convergent. We think this is probably due to the extremely simple spacetime structure induced by the infinite curvature, and does not necessarily signal an inconsistency by itself. Perhaps this is a rare case of nontrivial ``perturbatively complete" string theory, which nevertheless still contains the  infinite towers of oscillator modes as in usual critical strings. In any case in this paper we do not consider nonperturbative effects, assuming they are either negligible or nonexistent here.

\section{A universal upper bound for the entropy}

Our physical interpretation gives rise to a natural definition of the entropy $S_m(g)$ of a BMN string $O_{-m,m}$ as a function of string coupling constant $g$ 
\begin{eqnarray} \label{entropy}
S_m(g) = - \sum_{n=-\infty}^{\infty} p_{m,n} \log(p_{m,n}) .
\end{eqnarray} 
Basically the same formula appears as the Shannon entropy in classical information theory, von Neumann entropy in quantum theory, and also thermodynamic Gibbs entropy. We should provide some physical interpretations. The Shannon entropy should be well defined for any probability distribution, and represents the information channel capacity of a signal source with such probability distribution. On the other hand, since we are considering a single BMN string, instead of a macroscopic system, the thermodynamic Gibbs entropy interpretation seems not appropriate in our context. 

The interpretation as the von Neumann entropy in quantum theory is somewhat tricky, as the BMN state is a pure state. As in quantum theory, a pure BMN string $| m\ket$ evolves to another pure state $\hat{U}(g) |m\ket$ by a unitary operator. An observer can choose to make a quantum measurement in any orthogonal basis of the Hilbert space, which will give rise to a probability distribution. However, as explained in our previous paper \cite{Huang:2019}, the BMN states form a preferred basis because they are on-shell mass eigenstates, and their transition amplitudes are naturally computed by the dual CFT, as in Eq. (\ref{newentry}). It is helpful to think about the BMN strings as different types of particles, instead of as different quantum states of the same particle.  So a natural observer would perform quantum measurement in the BMN basis. In quantum theory, a measurement  can be usually understood as an entanglement process where the observer becomes entangled with the measured quantum state in the measurement basis. In our case when we trace out the observer, we get a mixed state of BMN strings with the probability distribution (\ref{pmatrix}). In this sense our definition (\ref{entropy}) can be also interpreted as the von Neumann entropy after an observer performs a measurement in BMN basis without revealing the measurement result. 

It is well known that the entanglement entropy of a bipartite quantum state is defined as the von Neumann entropy of the reduced density matrix from tracing out one party. Entanglement entropy has been well studied in the context of holographic duality, due to the seminar paper of Ryu and Takayanagi \cite{RT}, which proposed to compute entanglement entropy in conformal field theory holographically in terms of minimal surfaces in AdS space. In this paper we focus on the BMN string without further discussions on the aspect of entanglement with an observer during quantum measurement.

So, in the remaining parts of the paper, we simply refer to (\ref{entropy}) as the entropy of a BMN string, with the understanding of its main interpretation as the Shannon entropy of the probability distribution (\ref{pmatrix}), and a possible interpretation as the von Neumann entropy if we allow an unknown observer (or environment) to perform a measurement, which results in decoherence of a pure BMN state. For free string theory $g=0$, the probability matrix (\ref{pmatrix}) is simply an identity matrix, and the entropy for any mode $m$ vanishes $S_m(0) = 0 $. For finite string coupling $g$, our definition of entropy (\ref{entropy}) should be an intrinsic physical property of the BMN string, and it is sensible to study some of its salient features.

The zero modes represent discretized momenta in one of the eight traverse directions corresponding to the scalar insertion in BMN operators, while the positive and negative modes represent the left and right moving stringy excited modes. The total positive modes must cancel the negative modes due to the close string level matching condition \cite{BMN}. The conservation of (discrete) momentum in the traverse directions implies that $\bra \bar{O}_{0,0}^J O_{-n,n}^J \ket_{h} = 0$ for $n\neq 0$ at any genus $h$, which can be also directly confirmed by an integral formula like in (\ref{torusintegral}).  We should note a subtlety of the arguments here. As we mentioned in \cite{Huang:2019}, the non-negative condition of our probability interpretation only works for amplitudes with external single string states, while multistring states should be viewed as some kind of virtual states in the intermediate steps of a quantum process. Likewise, the ``conservation of zero mode" may be violated by multistring states, e.g. in the three vertex $ \langle\bar{O}^J_{-m,m}O^{J_1}_{0}O^{J-J_1}_{0}\rangle \neq 0$.

Summing over all genera in (\ref{pmatrix}), we have $p_{0,0}=1$ and $p_{0,n}=0$ for $n\neq 0$ for any coupling $g$. The zero mode BMN string is decoupled from the other modes, and the  entropy is simply $S_0(g)=0$. Here the formulas are symmetric for $\pm m$, so without loss of generality we can just from now on focus on $m>0$. 

For a quantum system with Hilbert space of finite dimension $D$, the maximal von Neumann entropy $\log(D)$  is achieved by a mixed state with uniformly distributed probability over an orthogonal basis. Since there are infinitely many BMN strings, it is not immediately clear that our  entropy (\ref{entropy}) is even finite. We shall prove an upper bound for the  entropy. 

For genus $h$, the field theory calculations of the two-point amplitude $\bra \bar{O}_{-m,m}^J O_{-n,n}^J \ket_{h}$ consist of $\frac{(4h-1)!!}{2h+1}$ cyclically inequivalent diagrams of dividing the long string into $4h$ segments \cite{HZ}. Due to cyclicity we only need to do a one-segment integral for one mode, and $4h$ integrals for the other mode. See e.g. the case of genus one (torus) in Eq. (\ref{torusintegral}). For $m\neq n$, the absolute value of a segment integration over a stringy oscillator mode is less than $\frac{1}{\pi |m-n|}$. So we have an upper bound for the two-point function 
\begin{eqnarray}
&& \bra \bar{O}_{-m,m}^J O_{-n,n}^J \ket_{h} \nonumber \\  &\leq&  \frac{(4h-1)!!}{2h+1} \frac{4h  g^{2h}}{\pi^2 (m-n)^2}    \int_{0}^1 d x_1 \cdots dx_{4h} \delta(\sum_{i=1}^{4h} x_i-1) 
\nonumber \\ &=& \frac{16h^2  g^{2h}}{2^{2h}(2h+1)!  \pi^2 (m-n)^2  }. 
\end{eqnarray}
We see that at large distance between mode numbers $|m-n|\sim \infty$, the strength of BMN string interactions are bounded by an inverse square law. Summing over all genera, we have an estimate of the probability matrix element (\ref{pmatrix}) as 
\begin{eqnarray}
p_{m,n} \leq \frac{f(g) }{\pi^2 (m-n)^2 },   \label{pbound}
\end{eqnarray}
where we denote a function which appears in the resulting summation as
\begin{eqnarray}
f(g):= \frac{2g }{ \sinh(g/2)}   [\frac{g^2+4}{2g} \sinh(\frac{g}{2})-\cosh(\frac{g}{2}) ]. 
\end{eqnarray}
The function goes like $f(g)\sim g^2$ in the large $g$ limit.

We can then estimate the  entropy (\ref{entropy}). First notice for $0<p<1$, the function $-p\log(p)$ achieved maximum at $p=e^{-1}$, and it is monotonic in $p\in(0,e^{-1})$. We can choose an integer 
\begin{eqnarray} \label{rangen0}
n_0 \geq \textrm{max}(\frac{\sqrt{e\cdot f(g) }}{\pi}, 2),
\end{eqnarray}
and evaluate the sum in three parts for $n\leq m-n_0$, $m-n_0< n< m+n_0$, and $n\geq m+n_0$. The two parts that extend to $\pm\infty$ are symmetric with the same contributions, and in the middle part the entropy is maximal with a uniformly distributed probability ensemble. We find 
\begin{eqnarray} \label{bound6}
S_m(g) \leq 2 \sum_{n=n_0}^{+\infty} \frac{f(g)}{\pi^2n^2} \log(\frac{\pi^2n^2} {f(g)})  + \log(2n_0-1) . 
\end{eqnarray}  
The sum is clearly convergent, so we get an upper bound for $S_m(g)$, which is actually independent of the string mode $m$. For large $n_0\sim \infty$, the infinite sum is infinitesimally small, and the dominant contribution comes from the second term $\log(2n_0+1)$.  To find the optimal upper bound, we look at the difference of the right-hand side of (\ref{bound6}) at $n_0+1$ and at $n_0$
\begin{eqnarray}  \label{diff}
\log(\frac{2n_0+1}{2n_0-1}) - 2  \frac{f(g)}{\pi^2 n_0 ^2} \log(\frac{\pi^2n_0 ^2} {f(g)}). 
\end{eqnarray}
The above expression is positive for large $n_0$, so we get better bounds when we decrease $n_0$ from infinity. 
The difference (\ref{diff}) as a function of $n_0$ may cross zero multiple times. To get a minimal value on the right-hand side of (\ref{bound6}), we need to check the integers $n_0$ where the expression (\ref{diff}) is positive at $n_0$ and nonpositive at $n_0-1$, in the range (\ref{rangen0}) up to a sufficiently large value.

String dynamics is usually quite difficult to analyze in the strong coupling limit, with notable exceptions due to many  revolutionary strong-weak dualities discovered in the 1990s, see e.g. \cite{Witten:1995}. Here due to the availability of convergent results up to all string loops, we can extrapolate to strong string coupling limit $g\rightarrow \infty$ and analyze the asymptotic behavior. In this limit the optimal upper bound (\ref{bound6}) is simply obtained by choosing the integer $n_0\sim g^{2+\epsilon}$, with an infinitesimal positive parameter $\epsilon$. For this choice the infinite sum in (\ref{bound6}) barely becomes infinitesimal. For sufficiently large coupling constant $g$, we can write a simple bound 
\begin{eqnarray} \label{logbound}
S_m(g) < (2+\epsilon) \log(g) , ~~~ g\sim \infty .
\end{eqnarray}
Usually, the maximal von Neumann entropy is interpreted as the logarithm of the ``effective dimension" of the Hilbert space. So we arrive at an interesting conclusion that although there are infinitely many BMN strings, the effective dimension is actually finite and grows at most a little more than quadratically as $g^{2+\epsilon}$ with the string coupling $g$.

Our analysis can be further applied to BMN strings with more stringy modes. The next simplest example is the BMN operator with three different stringy modes
\begin{eqnarray}  
O^{J}_{(m_1,m_2,m_3)} &=& \frac{1}{\sqrt{N^{J+2}}J} \sum_{l_1, l_2=0}^{J-1}  e^{\frac{2\pi im_2l_1}{J}} e^{\frac{2\pi im_3l_2}{J}} \nonumber \\ &&  \times \Tr(\phi^1 Z^{l_1} \phi^2 Z^{l_2-l_1} \phi^3 Z^{J-l_2}),  \label{operator3}
\end{eqnarray}
with the closed string level matching condition $m_1+m_2+m_3=0$. The probability amplitude and  entropy are defined similarly as before. There is a similar upper bound for the two-point function. For the generic case $m_i\neq n_i $, $i=1,2,3$, we have  
\begin{eqnarray} \label{logbound2}
 && \bra \bar{O}^{J}_{(m_1,m_2,m_3)} O^{J}_{(n_1,n_2,n_3)}  \ket_{h} \nonumber \\ &\leq & 
 \frac{64h^3  g^{2h}}{2^{2h}(2h+1)!  \pi^3 \prod_{i=1}^3 |m_i-n_i|}  . 
\end{eqnarray}
In the strong coupling limit, the upper bound for probability amplitude scales like $g^3$ instead of $g^2$ in (\ref{pbound}). We also dissect the summation range for the  entropy where now the dominant middle part is a two-dimensional domain bounded by $g^{3+\epsilon}$. Skipping the details, we derive 
\begin{eqnarray} 
S_{(m_1,m_2,m_3)} (g) < (6+\epsilon) \log(g) , ~~~ g\sim \infty .
\end{eqnarray}
In general we expect a universal logarithmic upper bound for  entropy with larger coefficients for BMN strings with more oscillator modes.

\section{Some numerical analysis}

In this section, we perform some numerical analysis to learn more about the  entropy of BMN strings.  In our previous paper \cite{Huang:2010}, we have computed the two-point function $ \bra \bar{O}_{-m,m}^J O_{-n,n}^J \ket_{h}$ up to genus $h=3$. We can use the vacuum correlator $ \bra \bar{O}^J O^J \ket_{h} = \frac{g^{2h}}{2^{2h} (2h+1)!}$ as a gauge of the numerical accuracy of the weak coupling approximation. For example, keeping up to $h\leq 3$ contributions, we get $99.8\%$ of the total contributions  of the vacuum correlators for $g=5$, or $90.8\%$ for $g=10$. So we expect our available data are good for some precise analysis up to $g\leq 5$, and for some rough analysis up to $g\leq 10$.

\begin{figure}
  \begin{center}
  \includegraphics[width=3.4in]{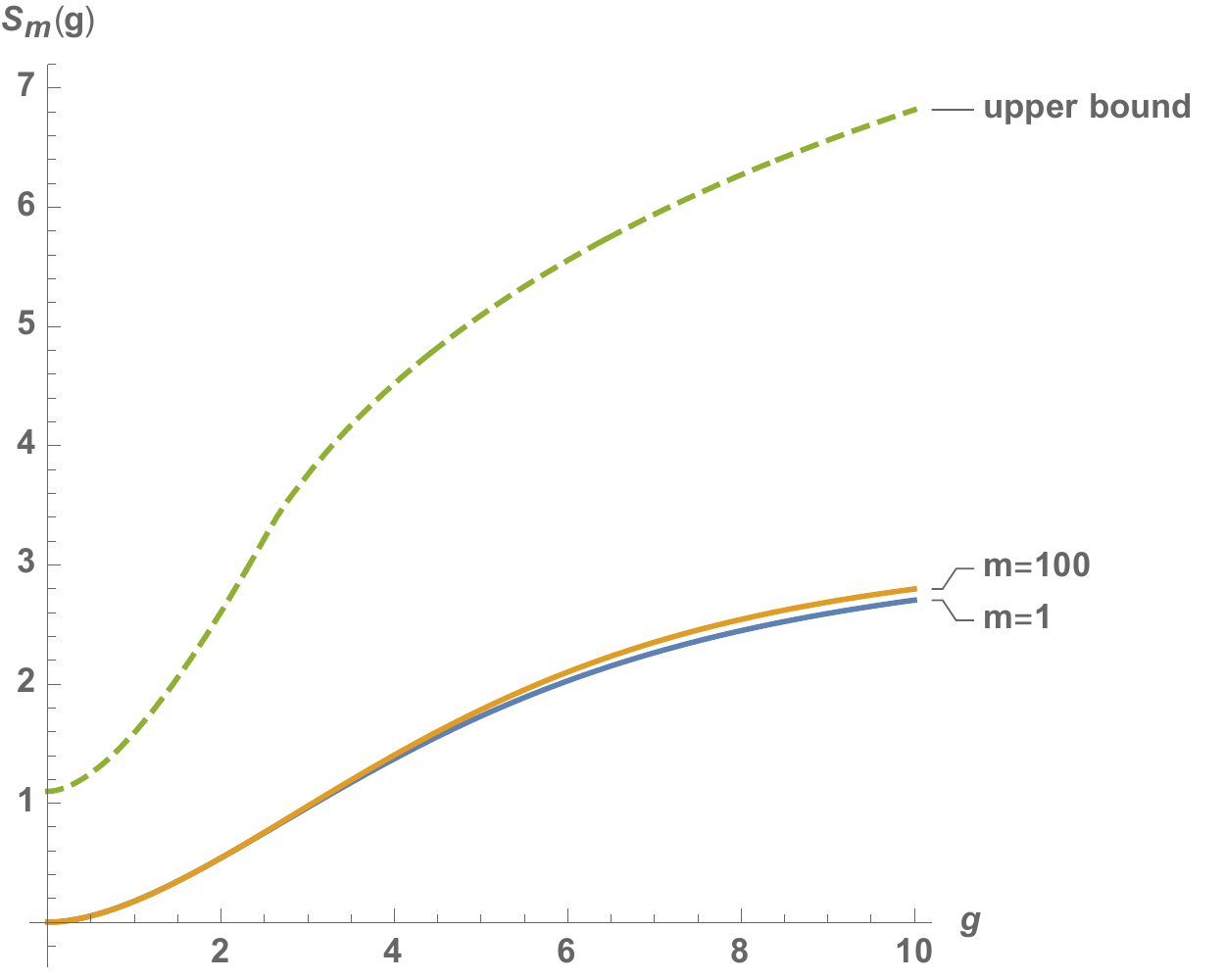} 
 \end{center}
\caption{The  entropy $S_m(g)$ for $m=1, 100$ and $0<g<10$. The dashed line is the upper bound (\ref{bound6}) with the choice of optimal integer $n_0$ in the range (\ref{rangen0}). }  \label{f1}
\end{figure}

As an illustration example we plot the  entropy $S_m(g)$ for two cases $m=1, 100$ and $0<g<10$ in Fig. \ref{f1}. We use the data up to genus 3 and also truncate the sum in (\ref{entropy}) at $|n|<10000$. The numerical accuracy is sufficient for our purpose. We see that the plots of actual values are consistent with and not too far off the upper bound derived analytically in (\ref{bound6}). 

We discuss two other salient features of the plot. First, for a fixed mode $m$, the function $S_m(g)$ appears to be a monotonically increasing function of $g$. This is  intuitively easy to understand. A BMN string $O_{-m,m}^J$ generally has stronger interactions with another string  $O_{-n,n}^J$ of nearby modes, i.e. smaller $|m-n|$, than those with far-off modes.   As the string coupling constant $g$ increases, the interactions have longer range and are more evenly distributed among strings with far-off modes, so the  entropy should increase. We have checked numerically many other examples that this monotonicity seems to be universally true at least for weak coupling $g$. However, it seems difficult to give a rigorous analytic proof. 

A second feature is that for a fixed string coupling $g$, the function $S_m(g)$ seems to depends very weakly on the string mode $m$. In Fig. \ref{f1} we see the plots for $m=1$ and $m=100$ are only slightly distinguishable. The function $S_m(g)$ for fixed $g$ has larger fluctuations around small modes, e.g. $m\leq 5$. However, as the mode number increases, the fluctuation becomes much smaller. The dependence on string mode $m$ is not monotonic. We check many examples that this is true at least for weak coupling. The intuitive explanation is that when we compute the two-point function as in the integrals in (\ref{torusintegral}), the dominant contributions come from the difference between mode numbers which appears in the integrals over stringy oscillator modes of e.g. $y_1,y_2$ in (\ref{torusintegral}), while the absolute mode number just contributes some oscillatory phases that give small fluctuations to the  entropy.
  
\section{Conclusion and Further Questions}  

We have studied the  entropy of BMN strings by analytic and numerical methods. It would be interesting to further improve the results. In particular, is our logarithmic bound (\ref{logbound}) close to optimal, or can it be much improved? For example, the  entropy may actually turn out to have a finite upper bound in the strong string coupling limit. We think this is not likely but cannot rule out this possibility. A more elaborate analysis is needed to answer these questions. 

Another interesting research direction is to explore whether there are some kinds of geometric interpretations of our results, or some interesting connections to entropy in other contexts. Motivated by the area law of black hole entropy, Bekenstein proposed a universal upper bound on entropy for bounded systems \cite{Bekenstein:1980}. The Bekenstein bound was later improved and generalized in many contexts. It would be interesting to explore whether our bound e.g. (\ref{logbound}) is somewhat related to this line of work. 

\

\noindent \textit{Acknowledgements}: This work was supported by the national Natural Science Foundation of China (Grants No. 11675167 and No. 11947301) and the national ``Young Thousand People" program. 

\

\end{document}